\documentclass[prl,twocolumn,showpacs,nobalancelastpage,amsmath,amssymb]{revtex4}
\usepackage{graphicx}
\usepackage{bm}
\begin{document}
\date{\today}

\title{
Quantum interference of edge supercurrents in a two-dimensional topological insulator  
}

\author{G. Tkachov, P. Burset, B. Trauzettel, and E.M. Hankiewicz}

\affiliation{
Institute for Theoretical Physics and Astrophysics, University of W\"urzburg, Am Hubland, 97074 W\"urzburg, Germany}

\begin{abstract}
Josephson weak links made of two-dimensional topological insulators (TIs) exhibit 
magnetic oscillations of the supercurrent that are reminiscent of those in superconducting quantum interference devices (SQUIDs). 
We propose a microscopic theory of this effect that goes beyond the approaches based on the standard SQUID theory.    
For long junctions we find a temperature-driven crossover from $\Phi_0$-periodic SQUID-like oscillations to a 
$2\Phi_0$-quasiperiodic interference pattern with different peaks at even and odd values of the magnetic flux quantum $\Phi_0=ch/2e$. 
This behavior is absent in short junctions where the main interference signal occurs at zero magnetic field.  
Both types of interference patterns reveal gapless (protected) Andreev bound states. 
We show, however, that the usual sawtooth current-flux relationship is profoundly modified by a Doppler-like effect of the shielding current 
which has been overlooked previously. Our findings may explain recently observed even-odd interference patterns 
in InAs/GaSb-based TI Josephson junctions and uncover unexplored operation regimes of nano-SQUIDs.
\end{abstract}

\maketitle

\begin{figure}[b]
\includegraphics[width=55mm]{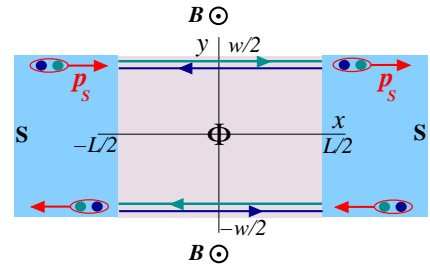}
\caption{(Color online) 2D TI with helical edge states between two superconductors (Ss)
in an out-of-plane magnetic field.
}
\label{SQUID_fig}
\end{figure}

{\em Introduction.}
In a two-dimensional topological insulator (2D TI) electric current flows near the edges of the system and is protected 
against elastic backscattering by time-reversal symmetry. 
This has important implications for both normal and superconducting transport in 2D TIs \cite{Hasan10,Qi11}. 
In very recent experiments \cite{Hart14,Pribiag14}, 2D TIs have been implemented as  
Josephson weak links between two superconductors. Remarkably, such Josephson junctions (JJs) act as
nanoscale superconducting quantum interference devices (SQUIDs) in which a magnetic flux, $\Phi$, 
enclosed in the interior of the 2D TI controls the interference of the Josephson currents flowing 
at the opposite edges of the sample (see also Fig. \ref{SQUID_fig}). 
The net supercurrent exhibits oscillations reminiscent of the SQUID pattern $\propto |\cos(\pi\Phi/\Phi_0)|$ 
rather than the Fraunhofer pattern observed in nontopological weak links. 

In view of the advances in fabricating 2D TI JJs and their application potential 
as topologically protected nano-SQUIDs, there is an apparent need for theoretical understanding of quantum interference phenomena in this type of JJs. 
This is the main motivation for our study.
One of the important questions is the following. 
The cosine-like SQUID pattern reflects the sinusoidal Josephson current-phase relation 
that can typically be attributed to the electronic states with energies close to or above the superconducting gap. 
However, the 2D TI JJs support also subgap Andreev bound states (ABSs) which are immune to non-magnetic disorder 
(see, e.g., Refs. \onlinecite{Fu09,Black-Schaffer11,Badiane11,Beenakker13,Crepin14}) and are 
highly anharmonic (e.g., sawtooth-like) with respect to the Josephson phase difference. 
One may therefore ask: How do such ABSs manifest themselves in a TI SQUID? 
We believe that the answer to that question may shed some light on the origin of the unusual "even-odd" interference patterns 
observed in InAs/GaSb-based TI JJs \cite{Pribiag14}.

To handle the problem we develop a nonperturbative treatment of the magnetic field effect on  
edge superconductivity which goes beyond the standard SQUID theory (see, e.g., Ref. \onlinecite{Tinkham96}) and more recent related studies \cite{Hui14,Baxevanis15}. 
The usual approach assumes that placing a JJ in a magnetic field does not perturb the superconducting contacts. 
In this approximation, the supercurrent depends only on the gauge-invariant Josephson phase difference determined by the flux $\Phi$ in the junction. 
While useful for conventional metallic JJs \cite{Tinkham96}, such a perturbative treatment 
faces severe limitations in hybrid structures where superconductivity is induced through the proximity effect 
and is suppressed already at a few millitesla \cite{Hart14,Pribiag14,Rohlfing09,Maier12}.    
Our theory lifts such limitations, showing a richer behavior of the Josephson current. 
We point out, in particular, the previously overlooked Doppler-like effect caused by the shielding response of the contacts to the applied magnetic field. 
This generic effect can be used to tune the ABS levels and detect them through the changes in the quantum interference patterns.

{\em Model.}
We consider a JJ created by depositing two $s$-wave superconducting (S) films on top of a 2D TI at a distance $L$ from each other (see Fig. \ref{SQUID_fig}). 
The width of the 2D TI, $w$, is assumed to be much larger than the total width of the two edge states. 
Then, each superconducting edge can be described by the quasi-one-dimensional Bogoliubov-de Gennes (BdG) Hamiltonian of the form
\begin{eqnarray}
&&
H_{BdG} = 
\left[
\begin{array}{cc}
 h(x) & i\sigma_y \Delta(x) e^{i\varphi_{_0}(x)} \\
 -i\sigma_y \Delta(x) e^{-i\varphi_{_0}(x)} & - h^*(x) 
\end{array}
\right], 
\label{H_BdG}\\
&&
h(x) =  v \sigma_x \, \left( -i\hbar\partial_x + \frac{p_{_S}}{2} \right) + U(x)- \mu.
\label{h}
\end{eqnarray}
Here $h(x)$ is the Hamiltonian for a given edge in the normal state,
$\sigma_x$ and $\sigma_y$ denote spin Pauli matrices, $v$ and $\mu$ are the edge velocity and Fermi energy, respectively. 
The potential $U(x)$ accounts for quasiparticle scattering in the JJ. The off-diagonal entries in $H_{BdG}$ 
incorporate a spin-singlet $s$-wave pair potential induced in the 2DTI underneath the S contacts. 
We assume that the superconducting gap $\Delta(x)$ and order-parameter phase $\varphi_{_0}(x)$ vary across the JJ as  
\begin{eqnarray}
\Delta(x) = \left\{
\begin{array}{cc}
 0, & |x| < \frac{L}{2}, \\
 \Delta, & |x| \geq \frac{L}{2}, 
\end{array}
\right. 
\quad 
\varphi_{_0}(x) = \left\{
\begin{array}{cc}
-\frac{\phi_{_0}}{2} , & x \leq - \frac{L}{2}, \\
+\frac{\phi_{_0}}{2}, & x \geq + \frac{L}{2}, 
\end{array}
\right.
\label{Delta_phi}
\end{eqnarray}
where $\phi_{_0}$ denotes the Josephson phase difference between the S regions in the absence of an external magnetic field. 
The magnetic field induces a local gradient of the superconducting phase, resulting in a finite Cooper-pair (condensate) momentum along the edge, $p_{_S}$, [see Fig. \ref{SQUID_fig} 
and Eq. (\ref{h})]. Using the gauge ${\bm A}(y) = (-By,0,0)$, one can relate $p_{_S}$ to the vector potential at the edge \cite{SOM}:
\begin{equation}
p_{_S}(B) = -\frac{2e}{c} A_x\left(\pm\frac{w}{2} \right) = \pm \, \pi\hbar \frac{Bw}{\Phi_0},
\label{p_S}
\end{equation}
where $\pm$ correspond to the upper ($u$) and lower ($l$) edges in Fig. \ref{SQUID_fig}, respectively. 

The Josephson currents for both upper and lower edges, $J_{u,l}$, can be obtained from the well-known scattering theory formula in the Matsubara representation 
(see, e.g., Refs. \onlinecite{Brouwer97,Dolcini07,Beenakker13}) 
\begin{equation} 
J_{u,l}(\phi_{_0},B) = -\frac{2e}{\hbar} k_BT \frac{\partial}{\partial \phi_{_0}} \sum_{n = 0}^\infty \ln D(\phi_{_0},B,\epsilon)|_{\epsilon = i\omega_n},
\label{J}
\end{equation}
where $\omega_n = (2n+1)\pi k_BT$ are the fermionic Matsubara frequencies for temperature $T$ ($k_B$ is the Boltzmann constant), and 
$D(\phi_{_0},B,\epsilon)$ is the characteristic function of energy $\epsilon$ whose zeros yield the energy spectrum of the JJ.  
Eq. (\ref{J}) assumes no constraints on the fermion parity, which is for instance the case for JJs strongly coupled to external reservoirs.
In Ref. \onlinecite{SOM}, we discuss how such constraints may affect our results. 
To find $D(\phi_{_0},B,\epsilon)$ explicitly, one should calculate the determinant of the system of eigenvalue equations 
obtained by matching the scattering states of $H_{BdG}$ at the boundaries $x=\pm L/2$ and at the scattering region \cite{Brouwer97}.    
We have done this calculation, taking into account the finite momentum $p_{_S}$ and assuming $\mu \gg \Delta, v|p_{_S}|$.  
The result is 
\begin{eqnarray} 
D(\phi_{_0},B,\epsilon) &=& \left(1-\alpha_{_<}  \alpha_{_>} {\rm e}^{i\frac{\beta_{_<} + \,\,\beta_{_>}}{2}} \right)^2  
\label{D1}\\
&-& {\cal T}\left(\alpha_{_<} {\rm e}^{i\frac{ \phi_0 + \beta_{_<} }{2}} - \alpha_{_>} {\rm e}^{-i\frac{ \phi_0 - \beta_{_>} }{2}} \right)^2.
\label{D2}
\end{eqnarray}
Here we use the subscripts $\gtrless$ to indicate the quasiparticle momentum direction -- {\em downstream} ($>$) or {\em upstream} ($<$) -- with respect to the condensate flow.  
In particular, $\alpha_{_\gtrless}$ are the amplitudes of Andreev reflection (AR) for the down- and upstream moving quasiparticles: 
\begin{equation} 
\alpha_{_\gtrless} = \frac{\Delta}{ \epsilon_{_\gtrless} +i\sqrt{\Delta^2 -\epsilon^2_{_\gtrless}} }, \qquad \epsilon_{_\gtrless} = \epsilon \mp \frac{v p_{_S}}{2},
\label{alpha}
\end{equation}
while $\beta_{_\gtrless}$ are the phase differences between particle and hole acquired in the normal region for the down- and upstream momentum directions:
\begin{equation} 
\beta_{_\gtrless} = \frac{2\epsilon_{_\gtrless}L}{\hbar v} = \frac{2\epsilon L}{\hbar v} \mp k_{_S}L, \qquad  k_{_S} = p_{_S}/\hbar.
\label{beta} 
\end{equation}
Importantly, $\alpha_{_>}\not = \alpha_{_<}$ and $\beta_{_>}\not = \beta_{_<}$ because the energies of the down- and upstream moving quasiparticles, $\epsilon_{_\gtrless}$, 
acquire opposite-sign shifts $\pm vp_{_S}/2$, a phenomenon analogous to the Doppler effect. It is a qualitatively new ingredient of the model, 
describing the coupling between the quasiparticles and superconducting condensate. For $p_{_S}=0$, Eqs. (\ref{D1}) -- (\ref{beta}) reproduce the corresponding results of Ref. \onlinecite{Beenakker13}. 
We note that the spin-momentum locking of the edge states guaranties their perfect transmission [${\cal T}=1$ in Eq. (\ref{D2})] for any potential $U(x)$ preserving time-reversal symmetry. 

Inserting Eqs. (\ref{D1}) -- (\ref{beta}) into Eq. (\ref{J}), we arrive at the following expressions for the Josephson currents at the upper and lower edges:
\begin{widetext}
\begin{eqnarray}
&&
J_u(\phi_{_0}, B) = -\frac{2e}{\hbar} k_BTi \sum_{n=0}^\infty 
\frac{
A_n^2(-B){\rm e}^{ i\phi } - A_n^2(B){\rm e}^{ -i\phi } 
}
{
[ 1 + A_n(-B)A_n(B)]^2 + [ A_n(-B){\rm e}^{ i\phi/2 } - A_n(B){\rm e}^{ -i\phi/2 } ]^2
}, 
\quad
J_l(\phi_{_0}, B) = J_u(\phi_{_0}, -B),\qquad
\label{J_u,l}
\end{eqnarray}
\end{widetext}
where the Josephson phase difference, $\phi$, includes the contribution of the external magnetic field,  
\begin{eqnarray}
\phi = \phi_{_0} + k_{_S}L = \phi_{_0} + \pi\Phi/\Phi_0,
\qquad 
\Phi = B L w,
\label{phi}
\end{eqnarray}
proportional to the flux, $\Phi$, piercing the area of the weak link, $Lw$, while the coefficients 
$A_n(B)$ absorb the AR amplitude together with the dynamical phase factor ${\rm e}^{i\epsilon L/\hbar v}$, 
both taken at imaginary energy $\epsilon=i\omega_n$:   
\begin{eqnarray}
A_n(B) = 
\frac{\Delta \, {\rm e}^{ -\omega_n L/\hbar v }
}
{
\omega_n + \frac{i}{2} v p_{_S}(B)  + \sqrt{ [\omega_n + \frac{i}{2} v p_{_S}(B)]^2 +\Delta^2  }
}. \,\,\,\,
\label{A}
\end{eqnarray}
The two edge currents differ only by the sign of the momentum $p_{_ S}$ [see Eq. (\ref{p_S})], 
which yields the second relation in Eq. (\ref{J_u,l}). 

Eqs. (\ref{J_u,l}) -- (\ref{A}) are our main results and merit a few comments here. 
As we can see, an external magnetic field has a two-fold effect. 
On the one hand, it generates oscillations of the current with the flux due to 
the extra Josephson phase difference $\pm \pi \Phi/\Phi_0$ 
induced at the upper ($+$) and lower ($-$) edges.  
Microscopically, this comes from the magnetic-field contribution $\mp k_{_S}L$ to the electron phase
in the weak link [see Eq. (\ref{beta})] and is expected on the general basis of gauge invariance. 
We can recast the total phase difference in the manifestly gauge-invariant form 
$\phi = \int^{_{L/2}}_{_{-L/2}} [\partial_x\varphi{_0}(x) - \frac{2e}{c\hbar}A_x\left(\pm \frac{w}{2}\right)]dx$. 
On the other hand, there is a shielding response of the S contacts to the applied field, causing Doppler-shifted AR with the reduced amplitude [see Eq. (\ref{A})]
when 
\begin{equation}
v|p_{_S}|/2 \geq {\max}(\Delta, \pi k_BT),  
\label{Doppler shift}
\end{equation}
or, explicitly,
\begin{equation}
B \geq B_{_{AR}}=\frac{2\Phi_0}{\pi w \xi_*}, \qquad   
\xi_* = {\min}\left( \frac{\hbar v}{\Delta}, \frac{\hbar v}{\pi k_BT} \right),
\label{Doppler field}
\end{equation}
where $B_{_{AR}}$ is the characteristic field for the AR suppression. 
This field scale competes with $B_{_{OSC}}=\Phi_0/\pi wL$ on which the oscillations occur, 
violating the periodicity of the currents $J_{u,l}$ with the magnetic flux (field). 
 
We wish to understand the interplay of the magnetic oscillations and Doppler effect in the Josephson current-field relationship which can be defined as   
\begin{equation}
J_m(B) = \left| J(\phi_{_0,max}, B) \right|, 
\label{Jm}
\end{equation}
where $J(\phi_{_0}, B) = J_u(\phi_{_0}, B) + J_l(\phi_{_0}, B)$ is the net current, and $\phi_{_0,max}$ is locked to the maximum current 
at zero field and given temperature, $J_m(0) = \left| J(\phi_{_0,max}, 0) \right|$. 
For the sinusoidal current-phase relationship (at $\pi k_BT \gg \hbar v/L$, see below), 
$J_m(B)$ is identical to the critical Josephson current (c.f. Ref. \onlinecite{Baxevanis15}). 
Generally, Eq. (\ref{Jm}) is less restrictive than that for the critical current because $\phi_{_0,max}$ fixes only the amplitude of the oscillations, 
allowing for a detailed analysis of their profile, which is important from both theoretical and practical points of view.
The notion of the critical Josephson current is an approximation itself that does not rule out the description based on Eq. (\ref{Jm}). 
In Ref. \onlinecite{SOM}, we prove that for our system $\phi_{_0}$ is gauge invariant and, therefore, can be fixed independently of the magnetic flux, 
i.e. Eq. (\ref{Jm}) satisfies the requirement of gauge invariance for the observables and is a valid description. 
In the following, we discuss $J_m(B)$ for long and short Josephson junctions.

\begin{figure}[t]
\includegraphics[width=65mm]{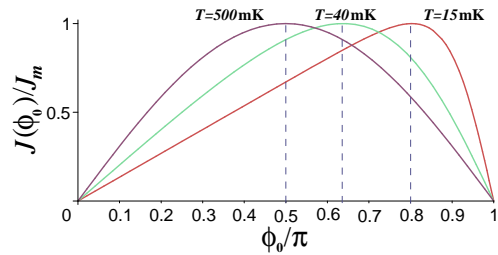}
\caption{(Color online) Zero-field current-phase relation $J(\phi_{_0})$ for a long junction with 
parameters close to those in experiments \cite{Pribiag14}:
$L=600$ nm, $\Delta=0.125$ meV, and $\hbar v=10$ meV$\cdot$nm. Curves for $T=15$, $40$ and $500$ mK represent the low 
($\pi k_BT \ll \hbar v/L$), intermediate ($\pi k_BT \sim \hbar v/L$) and high ($\pi k_BT \gg \hbar v/L$) temperature regimes. 
The phases $\phi_{_0,max}\approx 0.8\, \pi$, $\phi_{_0,max}\approx 0.63\, \pi$, and $\phi_{_0,max}\approx 0.5\, \pi$ 
yield the maximum current $J_m$ in each case.
}
\label{Long_phi_fig}
\end{figure}

{\em Long junction.}
It is a junction with the separation between the S terminals, $L$, much larger than the superconducting coherence length, $\xi=\hbar v/\Delta$.  
In the absence of the magnetic field, the junction current-phase relation has the well-known sawtooth shape at low temperatures $\pi k_BT \ll \hbar v/L$ \cite{Ishii70,Svidzinsky73}, 
turning sinusoidal in the opposite temperature limit $\pi k_BT \gg \hbar v/L$ (see Fig. \ref{Long_phi_fig}). 
For each temperature we determine the phase difference $\phi_{_0,max}$ yielding the maximum current $J_m$ and 
use that as an input for calculating $J_m(B)$ in Eq. (\ref{Jm}).

Figure \ref{Long_fig} shows the dependence $J_m(B)$ 
expressed in terms of the flux, $\Phi$, using $vp_{_S}/2=\pm (\pi \hbar v/2L) \, (\Phi/\Phi_0)$. 
At elevated temperatures (see Fig. \ref{Long_fig}a), 
we find SQUID-like oscillations due to the interference of two harmonic edge currents (cf. Fig. {\ref{Long_phi_fig}). 
As already mentioned, the current periodicity is broken by the superimposed  
reduction of the AR amplitude with increasing field.
For the same reason, the interference lobes decrease in magnitude when $B$ approaches the characteristic field $B_{_{AR}}$ (\ref{Doppler field}) 
which is considerably higher than $B_{_{OSC}}=\Phi_0/\pi wL$ for the chosen parameters. 
From Eqs. (\ref{J_u,l}) and (\ref{Jm}) with $\pi k_BT \gg \hbar v/L$, 
we derive an analytical formula for the SQUID-like current shown in Fig. \ref{Long_fig}a:     
\begin{eqnarray}
J_m(B) \approx \left|\,\frac{8ek_BT}{\hbar} \Re\left( A_0^2(B) e^{-i\pi \Phi/\Phi_0} \right)\,\right|.
\label{J_high}
\end{eqnarray}
Here $\Re$ means the real part and $A_0(B)$ is the zeroth-order coefficient from Eq. (\ref{A}). 
Eq. (\ref{J_high}) accounts for the magnetic-field dependence of $A_0(B)$ and is more general than the standard SQUID result 
$J_m(B) \propto \left|\, \cos(\pi \Phi/\Phi_0) \,\right|$. The latter follows from Eq. (\ref{J_high}) for $B \ll B_{_{AR}}$.

\begin{figure}[t]
\includegraphics[width=90mm]{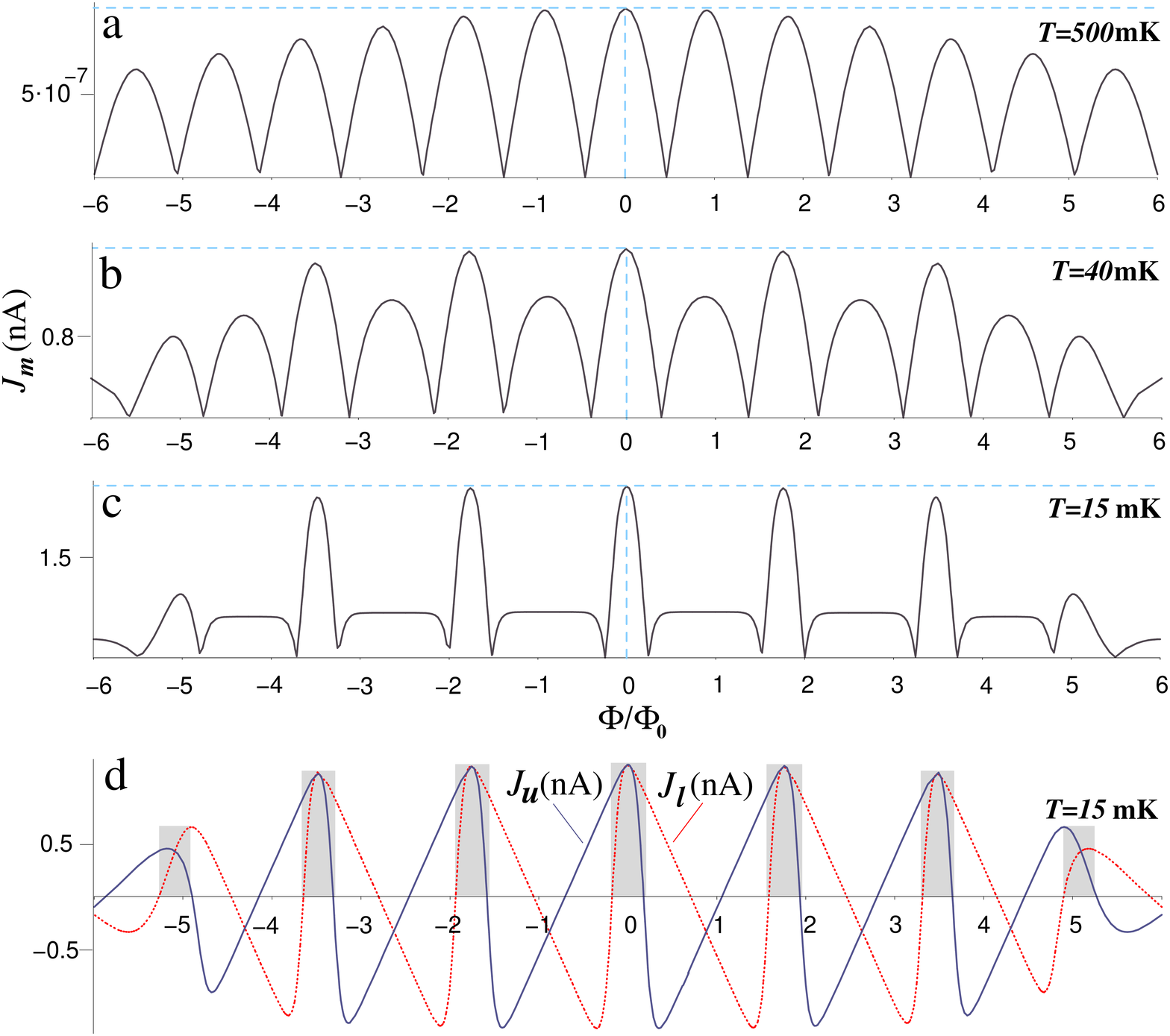}
\caption{(Color online) Supercurrent $J_m$ (\ref{Jm}) versus magnetic flux (panels a, b, and c) 
for a long junction with parameters specified in the caption to Fig. \ref{Long_phi_fig}. 
Panel d shows the current-flux relations for the upper and lower edges separately to explain 
the even-odd interference pattern shown in panel c.}
\label{Long_fig}
\end{figure}

With decreasing temperature, a $2\Phi_0$-quasiperiodic interference pattern emerges (see Figs. \ref{Long_fig}b and c).
This effect is most pronounced for $\pi k_BT \ll \hbar v/L$ when the current displays a series of alternating peaks with different heights and widths, 
as shown in Fig. \ref{Long_fig}c. With field-independent AR (i.e. for $B_{_{AR}}\to\infty$)
the alternating peaks would be centered exactly at the even and odd values of the flux quantum.
Such an even-odd pattern reflects the contribution of the ABSs. Their energies can be found from equation $D(\phi_{_0},B,\epsilon)=0$ (\ref{D1}) as
\begin{equation}
 \epsilon^\pm_k(\phi_{_0}, B) \approx 
 \frac{ \hbar v  }{2L}
 \left[ 2\pi k \pm \left(\phi_{_0} + \pi \frac{\Phi}{\Phi_0}\right)+2 \arccos\frac{ \mp vp_{_S}}{2\Delta} 
 \right],
\label{ABS}
\end{equation}
where $k=0,\pm 1,\pm 2...$ is an integer. This equation generalizes the well-known result of Ref. \onlinecite{Kulik70} by
taking into account the Doppler-shifted phases $\arccos(\mp vp_{_S}/2\Delta)$ from Andreev scattering off the moving condensate. 
For weak fields $vp_{_S}/2\Delta = B/B_{_{AR}}\ll 1$, 
the AR phases $\arccos(\mp vp_{_S}/2\Delta) \approx \pi/2$, and the ABSs 
are linear and $2\Phi_0$-periodic in flux. With increasing $B$, the field dependence of the AR phases distorts the sawtooth shape of the ABSs, 
which explains why the initially $2\Phi_0$-periodic pattern breaks down at higher fields where $B\gtrsim B_{_{AR}}$ [Fig. \ref{Long_fig}c]. 
Figure \ref{Long_fig}d shows that for each edge the current-flux relation initially has the sawtooth profile. 
The two currents tend to cancel each other except for the narrow regions 
near the current jumps (indicated by the shaded areas) where both currents flow in the same direction. 
In those regions $J_m(B)$ reaches the highest peaks becoming narrower with decreasing $T$ 
due to the increasing sharpness of the sawtooth pattern. 

\begin{figure}[t]
\includegraphics[width=55mm]{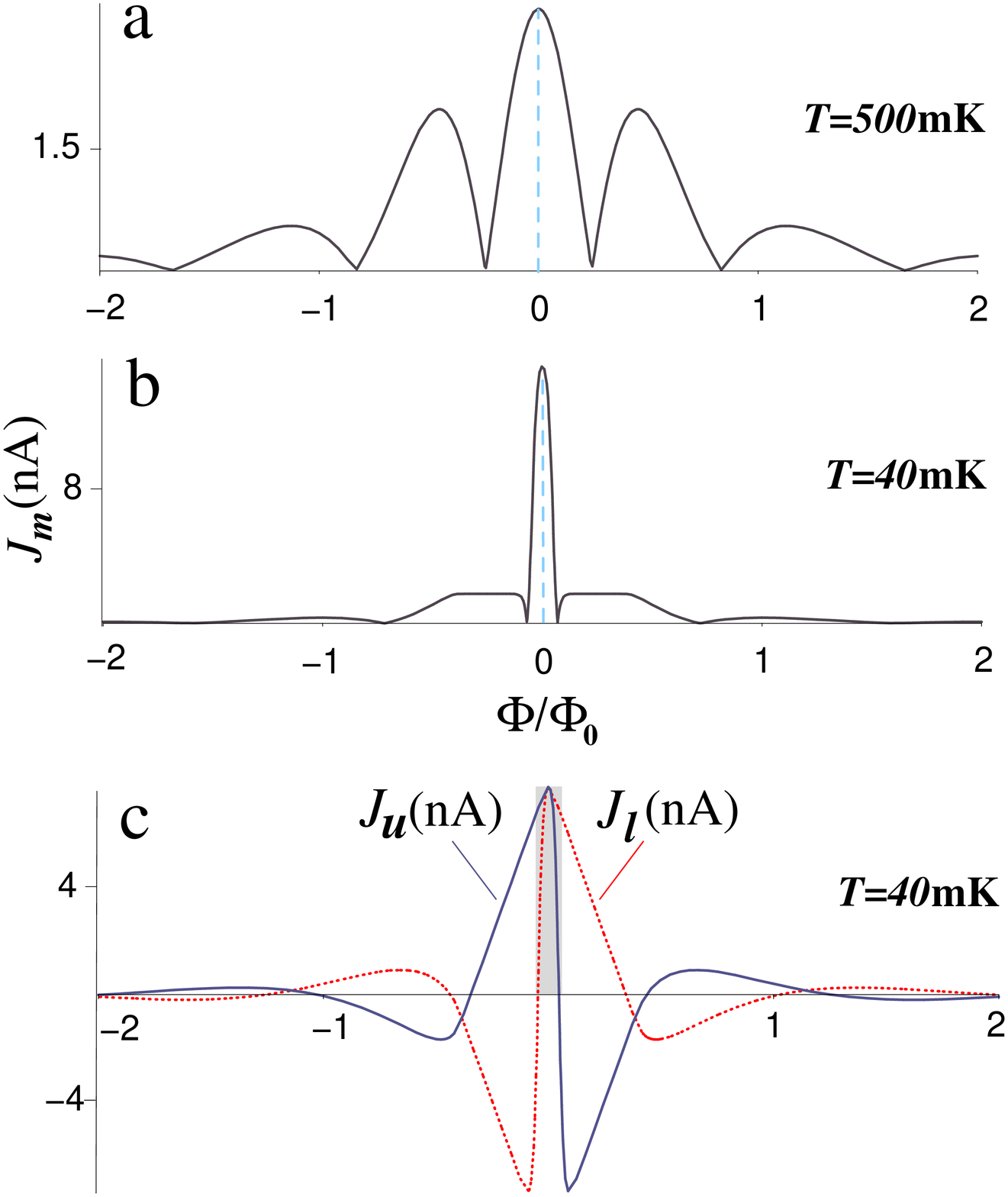}
\caption{(Color online) Supercurrent $J_m$ (\ref{Jm}) versus magnetic flux (panels a and b) for a short junction with 
$L=50$ nm, $\Delta=0.125$ meV, and $\hbar v=10$ meV$\cdot$nm. 
Panel c shows the current-flux relations for the upper and lower edges separately to explain 
the interference pattern shown in panel b.}
\label{Short_fig}
\end{figure}

{\em Short junction.} 
For short junctions with $L < \xi$ we predict a different type of quantum interference patterns.
In this case, the current oscillations are suppressed, showing no even-odd effect (see Fig. \ref{Short_fig}). 
This happens because the behavior of the current is dominated by the magnetic-field dependence of the Doppler-shifted AR.
Indeed, with decreasing junction length the field $B_{_{OSC}}=\Phi_0/\pi wL$ becomes comparable or even larger than $B_{_{AR}}$. 
For the interference pattern shown in Fig. \ref{Short_fig}a the fields $B_{_{OSC}}$ and $B_{_{AR}}$ are of the same order, 
while Fig. \ref{Short_fig}b corresponds to $B_{_{AR}} < B_{_{OSC}}$. In the latter case, 
AR is suppressed before the side interference lobes develop. The single zero-field peak in the low-T pattern 
is the direct consequence of the fact that a short junction supports only a single pair of ABSs below the gap.

{\em Conclusions.} 
We have studied the response of 2D TI JJs to an external magnetic field beyond the usual approximation 
where the Josephson current is determined only by the magnetic flux enclosed in the junction. 
In addition to this flux dependence, we have identified another magnetic-field effect that comes
from the shielding response of the contacts to the applied field and is characterized by Doppler-shifted Andreev reflection. 
Using such generalized approach, we have analyzed quantum interference of the edge supercurrents in both long and short JJs and 
identified the signatures of the topological ABSs.  
Our findings may have implications for interpreting recent experiments \cite{Hart14,Pribiag14}.
In particular, for Al-contacted InAs/GaSb quantum wells in the TI regime \cite{Knez}, Ref. \onlinecite{Pribiag14} reported SQUID-like oscillations 
disappearing at $B\approx 10$ mT. This observation agrees with our results shown in Fig.~\ref{Long_fig}a and 
with the estimate $B_{_{AR}}\approx 4$ mT for the suppression field (\ref{Doppler field}) 
for $3.9 \mu$m-wide JJs used in the experiment. With decreasing temperature, a crossover to 
the even-odd interference pattern was observed \cite{Pribiag14}.
This behavior is consistent with our predictions for the ABS contribution in a $600$ nm-long junction with $\Delta=0.125$ meV and 
$\hbar v=10$ meV$\cdot$nm \cite{Knez} for temperatures $T=40$ and $15$ mK  [see Figs. \ref{Long_fig}b and c].  
We also note that the measurement temperatures in Ref. \onlinecite{Pribiag14} correspond to a favorable regime for resolving individual ABSs because the 
resulting thermal smearing is smaller than the estimated level spacing $\hbar v/L \sim 0.05$ meV. 

\acknowledgments

This work was supported by the German Research Foundation (DFG Grant No TK60/1-1, FOR 1162, SPP1666, DFG-JST research unit "Topotronics") and 
the ENB graduate school "Topological insulators". We also thank Y. Ando, C. W. J. Beenakker, L. I. Glazman, D. Goldhaber-Gordon, T. M. Klapwijk, L. P. Kouwenhoven, 
A. Levy Yeyati, and L. W. Molenkamp for valuable discussions.

\newpage

\begin{appendix}
\section{Supplemental material to}
\begin{center}
{\bf "Quantum interference of edge supercurrents in a two-dimensional topological insulator"}

\vskip 0.25cm

G. Tkachov, P. Burset, B. Trauzettel, and E. M. Hankiewicz

\vskip 0.25cm

\textit{Institute for Theoretical Physics and Astrophysics, University of W\"urzburg, Am Hubland, 97074 W\"urzburg, Germany}
\end{center}

\subsection{Magnetostatics, symmetries, and gauge-invariant treatment of the system}

In this subsection, we discuss basic magnetostatic properties of the JJs considered in the main text. 
Using the symmetries of the magnetostatic problem, we determine the gauge-invariant Josephson phase difference under rather general assumptions about the screening 
properties of the system. We prove, in particular, the gauge invariance of the phase difference $\phi_{_0}$ [Eq. (11) of main text], 
which guaranties that the current $J_m(B)$ [Eq. (15) of main text] satisfies the requirement of gauge invariance for the observables.

\begin{figure}[b]
\includegraphics[width=85mm]{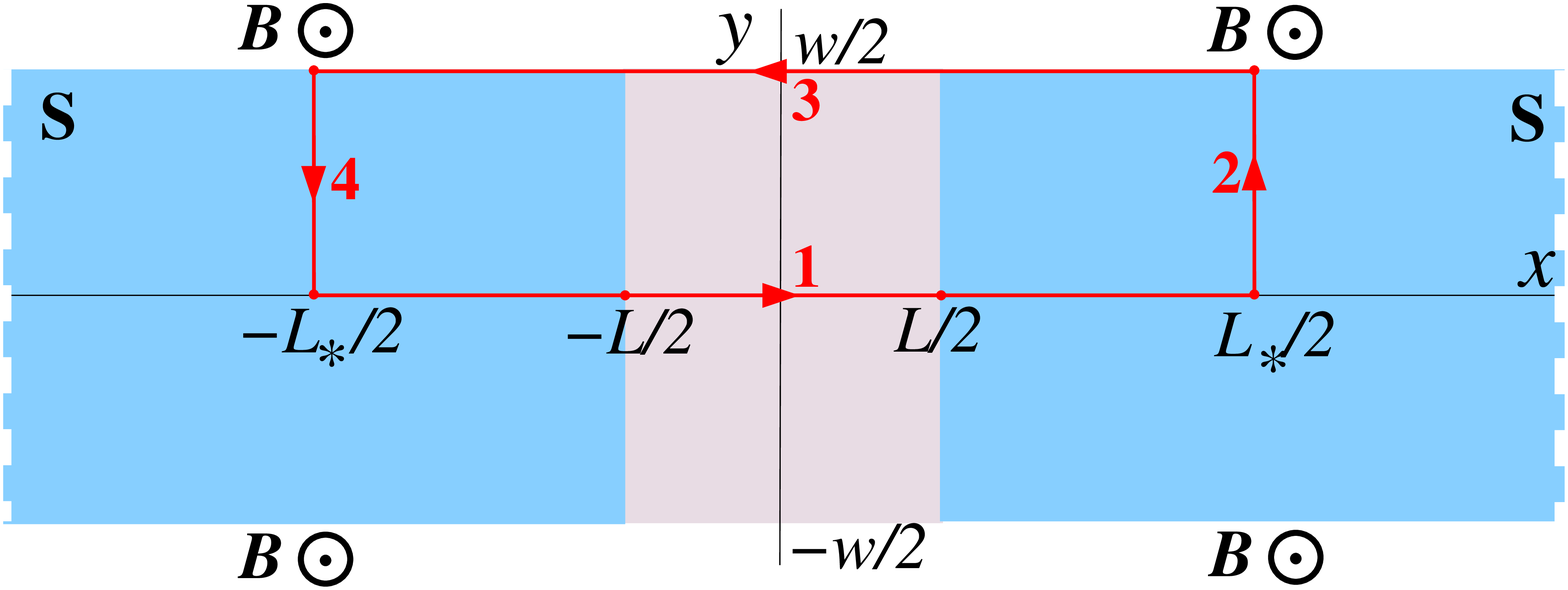}
\begin{flushleft}
FIG. S1: (Color online) Geometry of the system with a symmetrically applied external magnetic field.
Arrows show the integration path in Eq. (\ref{int}).
\end{flushleft}
\end{figure}

Let us discuss first general properties of the superconducting (S) leads in the presence of an external magnetic field ${\bm B}=[0,0,B]$ (see also Fig. S1). 
We consider semi-infinite leads and introduce the notation ${\cal {\bm B} }(x,y)=[0,0,{\cal B}(x,y)]$ for the magnetic field inside them. 
The latter is determined by Amp\`ere's law, which in Gaussian units reads 
\begin{eqnarray}
\partial_y\, {\cal B}(x,y) = \frac{4\pi}{c}j_x(x,y), \,\, -\partial_x \, {\cal B}(x,y) = \frac{4\pi}{c}j_y(x,y),
\label{Ampere}
\end{eqnarray}
where $j_x(x,y)$ and $j_y(x,y)$ are the components of the current density induced by the applied field.
The boundary conditions to Eqs. (\ref{Ampere}) are
\begin{equation}
{\cal B}(x,\pm w/2)=B,
\label{BC}
\end{equation}
and ${\cal B}(\pm L/2,y)=B$; the latter applies to the right and left boundaries, respectively. 
Note that the boundary conditions (\ref{BC}) imply the spatial symmetry ${\cal B}(x,-y)={\cal B}(x,y)$. 
Hence, the antisymmetric current density, $j_x(x,-y)=- j_x(x,y)$, vanishes in the middle of the superconductor, i.e.    
\begin{equation}
j_x(x,0)=0.
\label{j_x_0}
\end{equation}
For a type-II superconductor with the local relation ${\bf j} \propto \nabla \varphi$ between the current density and the gauge-invariant phase 
gradient \cite{Tinkham96},
\begin{equation}
\nabla \varphi = \nabla \varphi_0 -\frac{2\pi}{\Phi_0}{\bm A}, 
\label{grad}
\end{equation}
Eq. (\ref{j_x_0}) implies  
\begin{equation}
\partial_x \varphi(x,0)=0.
\label{grad_x_0}
\end{equation}
On the symmetry line $y=0$ within each superconductor the gauge-invariant phase $\varphi(x,0)$ remains constant and is equal to $\varphi(\pm L/2, 0)$. 

In order to determine the gauge-invariant Josephson phase difference we consider the line integral of Eq. (\ref{grad}) over the closed path shown in Fig. S1: 
\begin{eqnarray}
\oint \nabla \varphi d{\bm l} = 2\pi N - 2\pi\frac{\Phi/2}{\Phi_0},
\label{int}
\end{eqnarray}
where $N$ is the integer winding number of the order parameter phase $\varphi_0$, ensuring the uniqueness of the wave function, and 
we define the magnetic flux enclosed by the integration path as $\Phi/2$. 
The points $x=\pm L_*/2$ must be chosen away from the boundaries $x=\pm L/2$, where their influence on the current density is negligible, i.e. 
in the region where $j_y(x,y)\propto \partial_y\varphi(x,y) \approx 0$, and the integrals over paths 2 and 4 vanish. 
In view of Eq. (\ref{grad_x_0}), the integral over path 1 is equal to the phase difference across the junction, $\varphi(L/2, 0) - \varphi(-L/2, 0)\equiv \phi(0)$,
while the integration along path 3 yields the phase difference at the edge, $\varphi(-L_*/2, w/2)-\varphi(L_*/2, w/2)\equiv - \phi(w/2)$. 
Summing up all the contributions, we find       
\begin{eqnarray}
\phi(w/2) = \phi(0) + \pi \frac{\Phi}{\Phi_0} - 2\pi N.
\label{phi_+}
\end{eqnarray}
The same result with $\Phi \to -\Phi$ can be obtained for the edge at $y=-w/2$. 

We emphasize that the Josephson phase difference is related to the magnetic flux $\Phi$ through the {\em effective} junction area $L_*w$ 
which depends on the screening properties of the leads. 
In the main text, we assume, for simplicity, that the screening of the external magnetic field in the leads can be neglected, 
which is the main approximation in the small parameter $w/(2\lambda_P) \ll 1$, 
where $\lambda_P = 2\lambda^2_L/d_S$ is the Pearl penetration depth. 
It is the appropriate screening length for thin S films whose thickness in the $z$-direction, $d_S$, is much smaller than 
the London penetration depth $\lambda_L$. For $w/(2\lambda_P) \ll 1$ both $j_y$ and $j_x$ vanish in the main approximation, 
and we can use the linear vector potential ${\bm A}(y) = (-By,0,0)$ and choose $L_*=L$, which gives $\Phi = BLw$. 
Also, since $w$ is normally much larger than the transverse dimension of the edge states, 
the vector potential acting on the superconducting edge states can be approximated by its boundary value $A_x(\pm w/2)\approx \mp Bw/2$ [c.f. Eq. (4) of main text].    

Equation (\ref{phi_+}) proves the gauge invariance of the Josephson phase difference in Eq. (11) of main text where $\phi$ is a shorthand notation for $\phi(w/2)$. 
Furthermore, up to the unobservable phase difference of $2\pi N$, the constant $\phi_{_0}$ in Eq. (11) of main text 
coincides with the gauge-invariant phase difference in the middle of the junction, $\phi(0)$. 
Since the magnetic flux determines only the difference between $\phi(w/2)$ and $\phi(0)$, the gauge-invariant parameter $\phi_{_0}=\phi(0)$ can be fixed independently of $\Phi$.   
In our analysis, we choose $\phi_{_0}=\phi(0)$ to maximize in the gauge-invariant way the amplitude of the oscillatons of the Josephson current in Eq. (15) of main text. 

\subsection{Fermion parity effects}

The formula for the Josephson current, Eq. (5) of the main text, does not consider a fixed number of particles in the junction. 
Our results can be easily extended to consider the situation where the number of particles in the system is fixed and the ground-state fermion parity is conserved. 
Following Ref.~\onlinecite{Beenakker13}, the supercurrent at the upper and lower edges of this special junction is given by
\begin{equation}
 J^{u,l}_{\pm}=\frac{2e}{\hbar}\frac{\partial}{\partial\phi_0}F^{u,l}_{\pm} \quad ,
\end{equation}
with the free energies
\begin{eqnarray}
 F^{u,l}_0 &=& \frac{-1}{k_BT}\sum\limits_{n=0}^{\infty} \left.\ln D(\phi_0,B,\epsilon)\right|_{\epsilon=i\omega_n},  
\label{F_0}\\
 F^{u,l}_{\sigma} &=& F^{u,l}_0 -\frac{1}{k_BT}\ln\frac{1}{2} \left[ 1+\sigma(B)\mathrm{e}^{J_S}\sqrt{D(\phi_0,B,\epsilon=0)} \right. 
\nonumber \\
 &{}& \left. \times \exp \left(\sum\limits_{m=1}^{\infty}(-1)^m \left. D(\phi_0,B,\epsilon)\right|_{\epsilon=i\Omega_m}\right) \right],
\label{F_sigma} 
\end{eqnarray}
where $\Omega_m=m\pi k_BT$ are the bosonic Matsubara frequencies, and the function $D(\phi_0,B,\epsilon)$ is given by Eqs. (6) and (7) of main text. 
The $B$-field dependence is caused by the coupling of the quasiparticles to the condensate flow characterized by the momentum $p_S$, Eq. (4) of main text. 
Including this effect, the ground-state fermion parity is 
\begin{eqnarray}
 \sigma(B)&=& {\rm sgn} \left\{ 2\sqrt{1-\left(\frac{vp_S(B)}{2\Delta}\right)^2}\cos\left( \frac{\phi(B)}{2} \right) \right. 
\nonumber\\
 &{}& \left. - \frac{vp_S(B)}{\Delta} \sin\left( \frac{\phi(B)}{2} \right) \right\},
\label{sigma}
\end{eqnarray}
where $\phi=\phi_0+\pi\Phi/\Phi_0$ is the Josephson phase difference in a magnetic field. 
We set $\sigma=1$ when $\phi=0-k_S(B)L$ to solve the sign ambiguity. 
At the upper and lower edges, the parity-dependent supercurrent $J^{u,l}_{p}$ changes from $J^{u,l}_{+}$ to $J^{u,l}_{-}$ at $\phi=\pi\mp k_S(B)L$. 
Similarly to the junction without parity constraints, the currents at the edges are related by 
$J^l_p(\phi_0,B)=J^u_p(\phi_0,-B)$. The net Josephson current is thus given by $J_p(\phi_0,B)=J^u_p(\phi_0,B)+J^l_p(\phi_0,B)$. 
We focus on the long-junction case in which the lead-dependent factor $\mathrm{e}^{J_S}\sim 1$ 
(see Ref. \onlinecite{Beenakker13} for more details). 

\begin{figure}
	\includegraphics[width=0.9\columnwidth,angle=0]{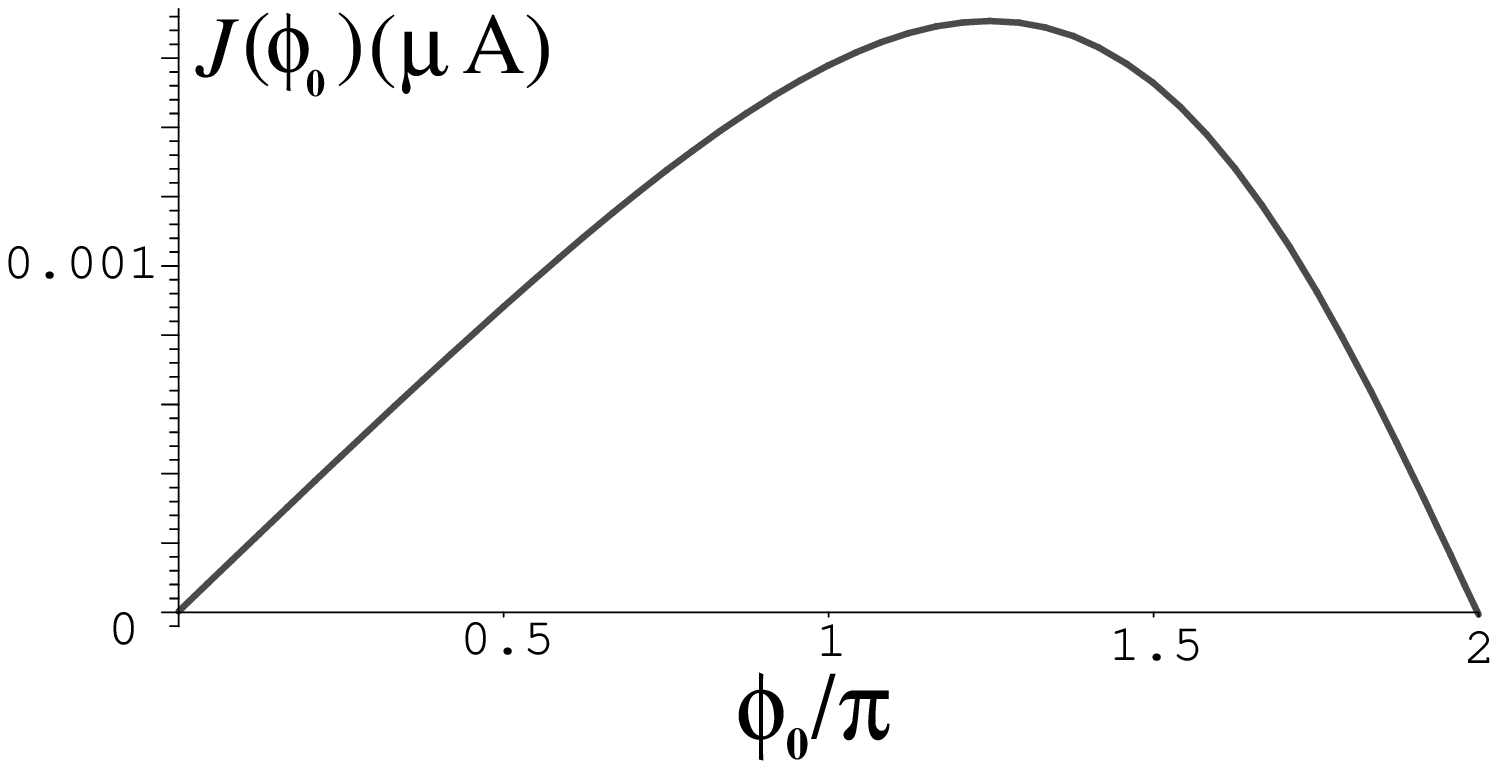}
\begin{flushleft}
FIG. S2: Parity-dependent current-phase relation for a long junction with the same parameters as in Fig. 2 of main text at $T=200$ mK.
\end{flushleft} 
\end{figure}


In Fig. S2 we plot the parity-dependent current as a function of the phase difference for the same junction parameters as in Fig. 2 of main text. 
The result looks similar to that in Fig. 2 of main text, but the period and amplitude of the current-phase relationship are doubled in agreement with Ref. \onlinecite{Beenakker13}. 
Consequently, the value of the phase $\phi_{0,max}$ locked to the maximum zero-field current is also doubled.


\begin{figure}[b]
	\includegraphics[width=1\columnwidth,angle=0]{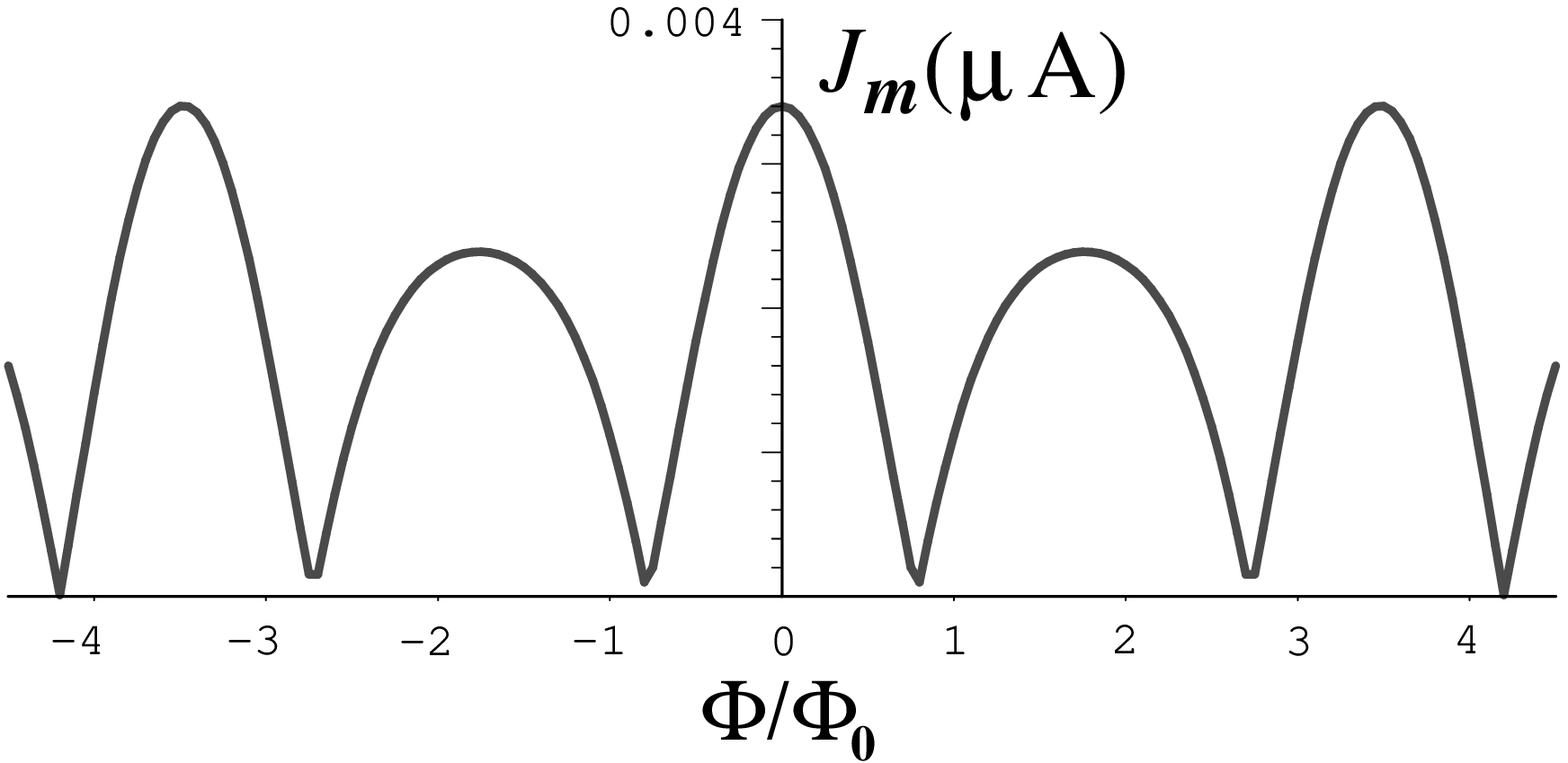}
\begin{flushleft}
FIG. S3: Parity-dependent supercurrent $J_m$ versus magnetic flux for a long junction with the same parameters as in Fig. 3 of main text at $T=200$ mK. 
\end{flushleft} 
\end{figure}


In Fig. S3 we show the flux dependence of the parity-preserving supercurrent. 
The results are qualitatively similar to those in Fig. 3b of main text, where no fermion-parity constraints were imposed, 
but with the period of the interference pattern equal to $4\Phi_0$ instead of $2\Phi_0$, a direct consequence of the $4\pi$-periodicity of the zero-field current.

\end{appendix}

\end{document}